# PROBABILISTIC FORECAST-BASED PORTFOLIO OPTIMIZATION OF ELECTRICITY DEMAND AT LOW AGGREGATION LEVELS

Jungyeon Park[1], Estêvão Alvarenga[2], Jooyoung Jeon[1], Ran Li[3], Fotios Petropoulos[4], Hokyun Kim[1] and Kwangwon Ahn[5]

1. Graduate School of Future Strategy, KAIST, South Korea
2. Shell, Netherlands
3. Department of Electrical Engineering, Shanghai Jiao Tong University, Shanghai, China
4. School of Management, University of Bath, UK
5. Department of Industrial Engineering, Yonsei University, South Korea

**Abstract**

In the effort to achieve carbon neutrality through a decentralized electricity market, accurate short-term load forecasting at low aggregation levels has become increasingly crucial for various market participants' strategies. Accurate probabilistic forecasts at low aggregation levels can improve peer-to-peer energy sharing, demand response, and the operation of reliable distribution networks. However, these applications require not only probabilistic demand forecasts, which involve quantification of the forecast uncertainty, but also determining which consumers to include in the aggregation to meet electricity supply at the forecast lead time. While research papers have been proposed on the supply side, no similar research has been conducted on the demand side. This paper presents a method for creating a portfolio that optimally aggregates demand for a given energy demand, minimizing forecast inaccuracy of overall low-level aggregation. Using probabilistic load forecasts produced by either ARMA-GARCH models or kernel density estimation (KDE), we propose three approaches to creating a portfolio of residential households' demand: Forecast Validated, Seasonal Residual, and Seasonal Similarity. An evaluation of probabilistic load forecasts demonstrates that all three approaches enhance the accuracy of forecasts produced by random portfolios, with the Seasonal Residual approach for Korea and Ireland outperforming the others in terms of both accuracy and computational efficiency.

*Highlights:*

· We proposed new low-voltage aggregation techniques based on portfolio theory to improve the accuracy of probabilistic electricity demand forecasting.

· The increased volatility at the low voltage level requires probabilistic forecasting to measure uncertainties.

· The proposed methods were found to outperform random aggregation with the real-world data from Korea and Ireland.

· A simple forecasting approach minimizing the standard deviation of deseasonalized demands was accurate and computationally efficient.

· The proposed method can contribute to achieving carbon neutrality, enable peer-to-peer energy sharing, and benefit policymakers, energy companies, and consumers.

# 1. Introduction

## 1.1. Motivation

The electric power industry is currently facing significant challenges in achieving carbon neutrality and security of supply, and two technological advances have offered new ways to address these challenges: (1) distributed energy resources (DERs) and (2) communications and control at the consumer level, including the use of smart meters and energy management systems [1]. With the growing penetration of DERs into the distribution grid, the power system is quickly evolving, necessitating the development of new retail market mechanisms for efficient investments and operations within a distribution system [2]. (1) New business models such as transactive energy, (2) grid optimization at all levels, and (3) DERs as providers of grid services are crucial elements for the future retail market in the context of DERs [3].

Hence, load forecasting at the lower aggregation level is becoming increasingly important for a variety of functions and the decision-making of market participants [4]. It is essential, for instance, for efficient peer-to-peer energy sharing, maintaining a balanced distribution grid, and utilizing demand response as a flexible resource. However, accurate load forecasting at low aggregation levels, such as a group of households, remains an issue, because load profiles aggregated to a consumer group are frequently more volatile and sensitive to individual behaviors than load profiles at high voltage (system) levels [4,5]. Furthermore, residential electricity load patterns have recently become varied among homes and neighborhoods due to shifting work and leisure patterns, increased use of electronics, and growing penetration of DERs such as rooftop photovoltaics, electric vehicles, and home-battery systems [6].

## 1.2. Related literature

There are an increasing number of studies on load forecasting [7,8]; however, most contributions are limited either to system-level loads or to individual household loads, which rely on smart meter data [4]. Recent studies on low-level aggregation load forecasting have focused on forecasting enhancement by (1) assessing and modifying traditional load forecasting techniques [9,10]; (2) building various models using machine learning, ensemble and hybrid methods, and probabilistic forecasting [11-20]; and (3) other methods such as data processing and clustering [4,9,21].

Since electricity usage may vary more at the home and building level than at aggregates, standard forecasting methodologies may not successfully adapt to the low aggregate level. The performance of various short-term load forecasting (STLF) techniques, from traditional linear regression to deep learning, is examined by Peng et al. [9] at various aggregation levels. The outcome reveals that all techniques perform better on residential loads with high aggregation than those with low aggregation levels. A number of STLF techniques currently in use, according to Dong-Ha et al. [10], are inappropriate for local load forecasting due to extended training time, instability in the optimization process, or hyper-parameter sensitivity. The modified double seasonal Holt-Winters model suggested in this paper performs reasonably well.

New forecasting models have been developed by utilizing machine learning, combinations of models, and probabilistic forecasting, which are emerging as research frontiers in the field of energy forecasting, to improve forecast accuracy at low aggregation levels.

The progress of machine learning and artificial intelligence techniques has certainly helped enhance energy forecasting, because identifying patterns or hidden information from historical data is critical for accurate forecasts [7]. A framework based on a long short-term memory (LSTM) recurrent neural network (RNN) is proposed by Kong et al. [11] to address STLF problems for residential households, which shows the best forecasting performance compared with multiple benchmarks. Voß et al. [12] demonstrate that convolutional neural networks (CNN) called WaveNet outperforms the benchmarks for the aggregation of 10~200 households.

Various ensemble and hybrid models have recently been described for improving forecasting accuracy ([4], [7]). Yang et al. [13] propose a dynamic ensemble method for residential STLF, employing a vector autoregression (VAR) model, Gaussian process regression (GPR) model, and LSTM neural network, which shows better performance than other state-of-the-art ensemble approaches. A technique that utilizes feature engineering, pooling, and a hybrid LSTM-SAM (self-attention mechanism) model is proposed by Zang et al. [14] to enhance the accuracy of day-ahead residential load forecasting.

Due to the increase in energy market uncertainty over the last decade, energy forecasting has shifted from a deterministic to a probabilistic perspective, which offers more information about future uncertainties than a point prediction [7]. Various probabilistic forecasting methods, such as quantiles, intervals, and density functions, have been suggested [15]. However, when compared to the literature on probabilistic forecasting in general or probabilistic wind power forecasting [16,17], the literature on probabilistic load forecasting is relatively sparse. Hong and Fan [8] recently reviewed load forecasting with an emphasis on the shift from point to probabilistic load forecasting. Conditional kernel density (CKD) estimation was used by Arora and Taylor [18] to obtain probability density estimates for power usage, which are then utilized to derive prediction intervals for electricity costs for various tariffs. Guo et al. [19] present a probability density forecasting approach based on deep learning, quantile regression, and kernel density estimation (KDE), and Jeon et al. [20] propose a probabilistic forecast reconciliation method to improve the prediction accuracy of power load.

In addition to forecasting models, research is developing a variety of different approaches to enhance prediction accuracy at lower aggregation levels. One option for reducing the forecasting error rather than utilizing sophisticated learning algorithms is to process the data to increase predictability [9]. In Bui et al. [21], a statistical data-filtering technique with the best confidence interval of the input dataset was developed to address the unexpected noises/outliers of the database to increase the accuracy of STLF in distribution networks. Load profile clustering can also be utilized to improve forecast accuracy by identifying customers' electricity consumption patterns. This is accomplished by first aggregating comparable groups, creating a prediction for each group using different models, and then adding the forecasts of the groups to produce the aggregated load forecast [4].

Although several recent research studies have shown that the suggested approaches enhance prediction performance, little work has been done to address the optimal portfolio of electricity demand, which could be utilized for developing the strategies of market participants as well as improving forecasting accuracy. Portfolio optimization, which seeks a portfolio that achieves high expected returns and minimized risks [22], has been used to deal with uncertainty in the energy sector, however, it is primarily focused on the supply side for the optimal renewable energy portfolio [23-25] rather than the electricity demand side.

### 1.3. Research objectives and contributions

In this paper, rather than seeking a better forecasting model, we propose a new methodology to construct an aggregated demand portfolio with minimal probabilistic forecast errors for a given energy demand, which can also drastically increase forecast accuracy. We analyze the relationship between the expected aggregated demand and density forecast errors, which are referred to as 'expected return' and 'risk', respectively, in the classical portfolio theory, as illustrated in Figure 1. There are numerous combinations of individual demand that create a specific aggregated demand on the Y axis, but the purpose of this study is to find a combination that minimizes the forecast inaccuracy on the X axis. The optimal electricity demand portfolio with high forecast accuracy for a particular energy demand at a certain lead time, known as the efficient frontier, can be used by market participants to promote efficient power system operation by effectively managing peer-to-peer (P2P) energy sharing and providing flexibility resources via demand response (DR) programs. This approach, however, differs from the classical portfolio theory in that it focuses on lowering risk for a given energy demand and does not account for portfolio weight, which implies a proportion of the consumer's demand.

Instead of deterministic forecasting, we generate probability density forecasting. Probabilistic demand forecasts at different locations in the grid are essential for optimum decision-making [26] because simple point forecasting methods, which calculate a conditional mean of future demand, cannot fully capture the forecast uncertainty, and therefore lead to unsatisfactory decision-making [27]. In other words, probability density forecasting can represent more uncertain information about future load fluctuation that will be more favorable for the decisions on the purchase and sale of electricity, the optimization of home energy management systems (HEMS), or modeling of the distribution grid [28-30]. Limited research has been conducted on stochastic load forecasting compared to wind power forecasting [17], but its prominence has shown considerable growth [8].

The aim of this paper is to identify the optimal method in terms of load forecast accuracy and computing efficiency to aid market participants' fast and accurate decision-making. To find the best approach, we introduce three different methods to assemble a portfolio of residential households' demand and evaluate the density forecasts of group demand up to a day ahead. We consider forecasting methods based on either the univariate autoregressive moving average generalized autoregressive conditional heteroskedasticity (ARMA-GARCH) model or an unconditional KDE model. We used the continuous ranked probability score (CRPS) and the mean absolute error (MAE) to assess the accuracy of the forecasts.

The main contributions of this study are summarized below:
(1) Novel low voltage aggregation methods based on the portfolio theory to minimize electricity demand forecasting uncertainties are proposed.
(2) Probabilistic load forecasting at low aggregation levels is used to capture and measure future uncertainties.
(3) The proposed methods are examined using real-world data from Korea and Ireland to demonstrate their forecasting performance and practical applicability.
(4) An optimal portfolio of electricity demand with minimal forecasting error can be utilized for strategic decision making by various market participants such as P2P operators, DR aggregators, and distribution system operators (DSOs).

The rest of this paper is organized as follows. Section 2 presents the smart meter data description and seasonal decomposition. Section 3 describes the ARMA-GARCH and KDE model for density forecasting and the CRPS and MAE as evaluation criteria. The forecast accuracy at various aggregation sizes and lead times is provided in Section 4. Section 5 describes forecast-based portfolio optimization, and Section 6 discusses the forecasting performance of the proposed methods and practical implications. Finally, the conclusion is presented in Section 7.

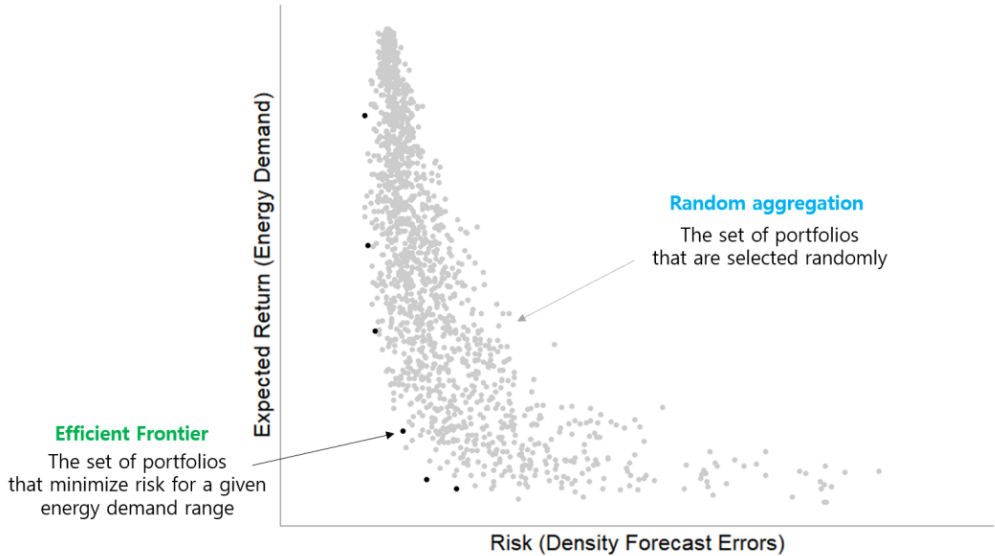

**Figure 1**. The expected demand for a forecast-based portfolio of consumers and its risk measured as the forecast error (dark grey points represent the efficient frontier).

**2. Data and decomposition**

*2.1. Smart meter data*

This study uses Korean and Irish smart meter data. The Korean dataset consists of an hourly electricity consumption series of 1,000 residential customers in Korea, recorded between January 2012 and October 2014: a confidential sample provided by the Korea Electric Power Corporation (KEPCO). The Irish smart meter data is publicly accessible from the Irish Commission for Energy Regulation (CER) [31]. Half-hourly data for the electricity consumption of 651 residential customers in Ireland, recorded between July 2009 and December 2010, are converted to hourly data.

Figure 2 illustrates the demand time series patterns of four different households over the same four-week period in February 2012 in Korea and February 2010 in Ireland. The data includes 24h and 168h (one week) cycles. The individual residential demand certainly shows high diversity and volatility.

The Korean data had an average of 3.5% missing values, while the Irish data had 0.03% missing values. To retain the daily and weekly seasonal patterns, the value of the same hour on the same day of the previous week was substituted for missing observations.

For the empirical analysis, we used multiple samples drawn from the data by using sliding window evaluation, as illustrated in Figure 3. Each sliding window consists of a train set and a test set. Following the completion of each training and evaluation of forecasting performance, the next sample is drawn from the data after a prime number interval. Rolling hourly samples for 12 weeks were used for model fitting and forecasting the next 72 hours to keep the computational cost minimal without seasonal bias.

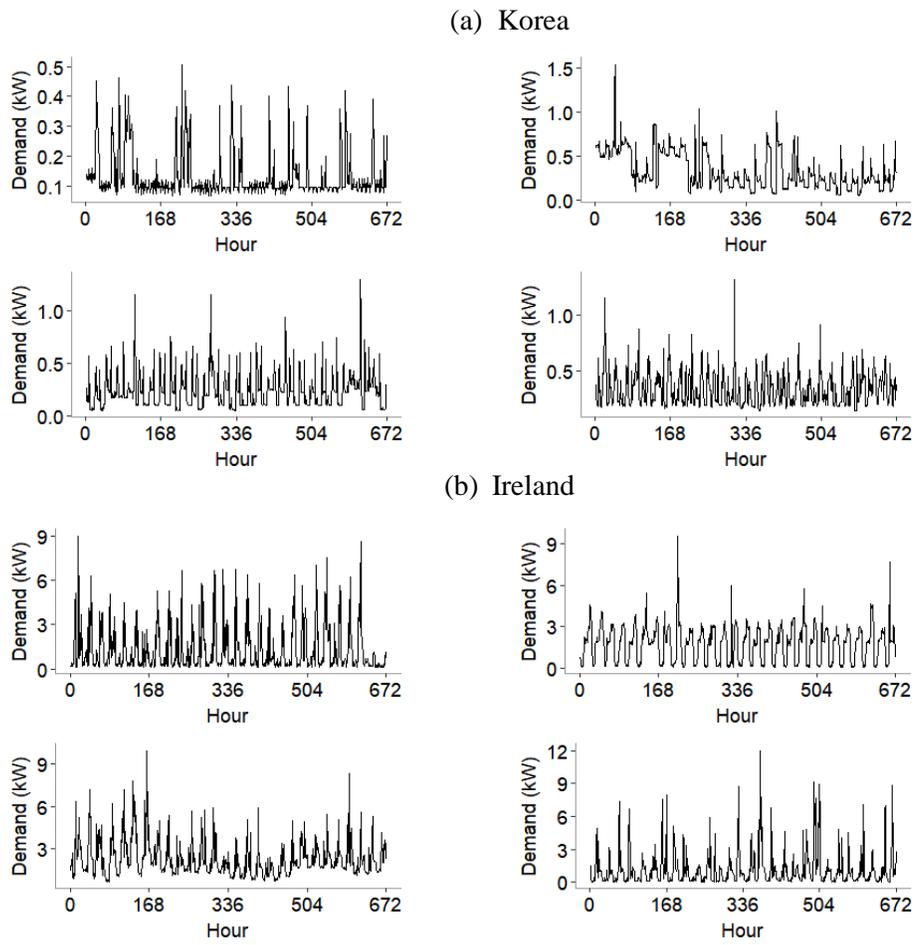

**Figure 2**. Example of electricity demand pattern for four different households during a four-week period.

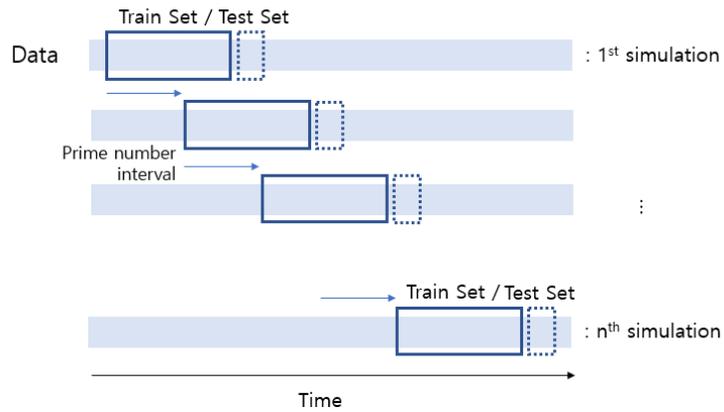

**Figure 3**. Sliding window evaluation with a prime number interval.

## 2.2. Decomposition

Electric load time series frequently contain several seasonal cycles (i.e., daily and weekly cycles), distinct characteristics that might impose unnecessarily high inferential or computational costs for developing a model [32]. Most load forecasting methods generate forecasts primarily by modeling the original time series, however, various seasonal patterns in the original data may increase the complexity of the forecasting model. As a result, extracting independent features from the original data provides an alternative to building separate predictors to improve forecast performance while also providing a suitable treatment of nonlinear characteristics [33].

The most traditional way of dealing with seasonality in time series is to use a seasonal decomposition algorithm like the X-11 method to extract it [34]. Aside from work on the X-11 technique and its variants, numerous other seasonal adjustment methods have been developed, one of which is the nonparametric Seasonal-Trend decomposition procedure based on Loess (STL) method proposed by Cleveland et al. [35]. Unlike the X11 decomposition technique, STL is widely used as it can handle missing information and deal with any kind of seasonality, not only monthly and quarterly data. It can also be resilient to outliers, ensuring that estimates of the trend-cycle and seasonal components are unaffected by a few odd observations [36].

STL is a filtering procedure for decomposing the original data $Y_t$ into seasonal $S_t$, trend-cycle $T_t$, and remainder $R_t$ in Equation (1), using a local regression technique known as Loess:

$$Y_t = S_t + T_t + R_t \tag{1}$$

In this paper, we deseasonalized the time series using STL as the first step of our procedure, as demonstrated in Figure 4.

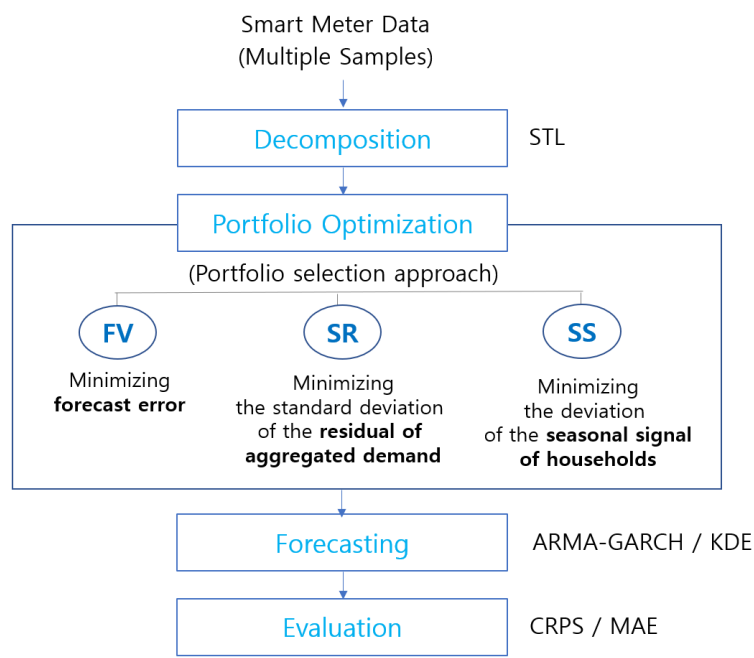

**Figure 4**. Framework of forecast-based portfolio optimization approaches.

## 3. Density forecasting

In this section, we describe the forecasting models and evaluation criteria required for portfolio optimization for demand aggregation in Section 5. While ARMA-GARCH models produce density forecasts by modeling the conditional mean and variance, a simpler approach to density forecasting is KDE, which summates kernel functions with a specific bandwidth [18]. We fit an ARMA-GARCH model to the time series of each consumer or each group of consumers in each time step to produce forecasts, as stated in Section 3.1. The adequacy of the model fit is assessed using a Pearson chi-squared goodness-of-fit test, as used by Vlaar and Palm [37]. If the obtained p-value of the test is lower than the predefined model threshold detailed in Section 3.3, the null hypothesis, that the observed frequency distribution in a sample is consistent with a theoretical distribution, is rejected. In this situation, we apply the KDE stated in Section 3.2 instead of the ARMA-GARCH model. To evaluate the probabilistic and deterministic forecast accuracy of the forecasting models, CRPS and MAE are used, respectively, as described in Section 3.4.

### 3.1. ARMA-GARCH

To capture autocorrelation in the conditional mean and variance, ARMA-GARCH models are extensively employed in the electricity industry. For wave energy forecasting, Jeon and Taylor [38] apply ARMA-GARCH models, a regression method, and conditional KDE. Liu and Shi [39] use various ARMA-GARCH models to model and forecast hourly ahead electricity prices from the New England electricity market. Garcia et al. [40] also present a GARCH model for forecasting hourly electricity prices in the deregulated power markets of Spain and California. Bikcora et al. [41] investigate the ARMA-GARCH model for day-ahead load forecasting at a local level.

In this paper, we use the ARMA $(p, q)$ - GARCH $(r, s)$ model for short-term load forecasting at low aggregation levels. The ARMA $(p, q)$ represents the conditional mean process, while the GARCH $(r, s)$ describes the conditional variance process. The GARCH $(r, s)$ process is similar to an ARMA process implemented to model the variance magnitude. The ARMA $(p, q)$ - GARCH $(r, s)$ model follows Equations (2a) to (2c),

$$y_t = \alpha_0 + \sum_{j=1}^{p} \alpha_j y_{t-j} + \sum_{k=1}^{q} \beta_k \varepsilon_{t-k} \tag{2a}$$

$$\sigma_t^2 = \phi_0 + \sum_{l=1}^{r} \phi_l \sigma_{t-l}^2 + \sum_{m=1}^{s} \gamma_m \varepsilon_{t-m}^2 \tag{2b}$$

$$\varepsilon_t = \sigma_t \eta_t, \tag{2c}$$

where $y_t$ is the energy demand observed at time $t$; $\varepsilon_t$ is an error term; $\sigma_t$ is the conditional standard deviation implying volatility; $\alpha$, $\beta$, $\phi$ and $\gamma$ are the coefficients of the AR, MA, GARCH and ARCH components with orders denoted by non-negative integers $p$, $q$, $r$ and $s$, respectively; and $\eta_t$ is the white noise generating process. In principle, the $\varepsilon_t$ could follow any suitable distribution model.

We use the R package "rugarch" implemented by Ghalanos [42] to build the ARMA-GARCH model. The ARMA $(p, q)$ order is chosen with the lowest Bayesian Information Criterion (BIC) defined by Schwarz [43]. The GARCH $(r, s)$ order is set to (1,1) which has shown to produce a simple and relatively accurate result for volatility estimation across many disciplines [44, 45]. We assume a skew generalized error distribution (SGED) to capture the skewness of demands, as shown in Figure 5.

We concentrate on the ARMA-GARCH model without comparing it to any other forecasting models, because the goal of this study is to identify the best aggregated demand portfolios with minimal prediction error rather than to develop a detailed forecasting model.

*3.2. Kernel density estimation*

KDE avoids any previous assumptions of distributions and generates the density function based on historical observations. We implement unconditional KDE, similar to those presented in Taylor and Jeon [17] for probabilistic wind forecasting, Arora and Taylor [18] for household-level probabilistic load forecasting and Jeon and Taylor [38] for short-term density forecasting of wave energy. This method produces an estimate of a probability distribution function $f(y)$ of the energy demand $y$ based on observations $\{y_1, y_2, \dots, y_n\}$. Unconditional KDE can be defined by Equation (3),

$$f_{KDE}(y|t) = \sum_{w=1}^{W} K_h(y - y_{t-w}),\tag{3}$$

where $y$ is the energy demand forecast to be estimated, $W$ is the number of observations to be included in kernel smoothing, and $K_h$ is a Gaussian kernel function with bandwidth $h$.

It is critical to adequately estimate the kernel bandwidth, because a large bandwidth leads to over-smoothing of the key underlying features of the time series, whereas a small bandwidth causes under-smoothing, resulting in an estimated density that is too rough [18]. The bandwidth in this paper is chosen according to Silverman's reference bandwidth, also known as Silverman's rule of thumb [46]. Figure 5 exemplifies density plots for four different households at the same 24 hours without prior seasonal analysis.

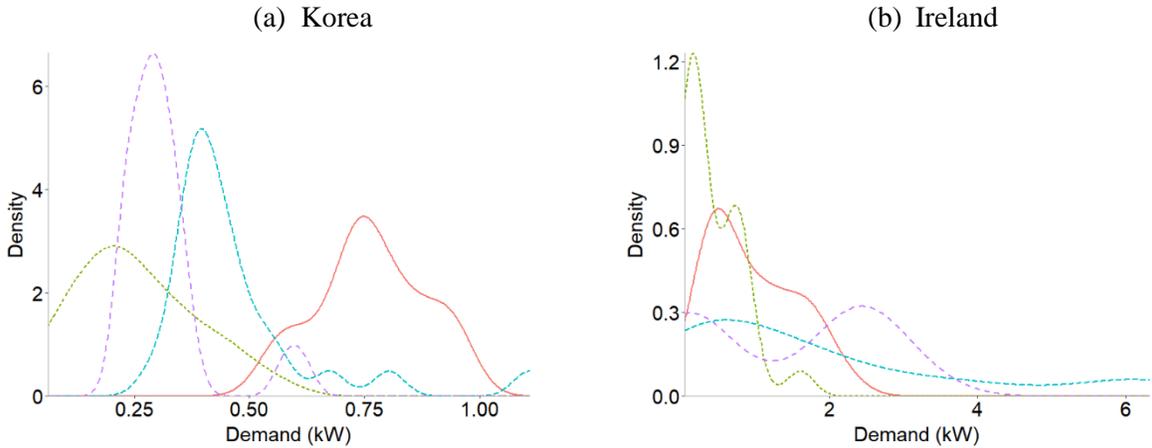

**Figure 5**. Kernel densities for four different individual households' demands estimated with the same 24-hours chosen within the in-sample.

*3.3. Model selection*

Although the ARMA-GARCH model is known to perform well in load forecasting, it might not be suitable for modeling certain consumption behavior. Therefore, we perform a Pearson chi-squared goodness-of-fit test in the "rugarch" package. This returns the p-values as well as a statistic of adjusted goodness-of-fit for the fitted distribution according to the study by Vlaar and Palm [40]. If the p-value is larger than a pre-defined threshold, ARMA-GARCH is used for forecast modeling. Otherwise, KDE is used.

To choose the optimal threshold $\delta$, we tried various threshold values ranging from 0 and 0.2 with increments of 0.01. The exploration was performed with individual households and randomly aggregated demands for the consecutive but distinctive 12-week fitting periods, followed by forecasts from 1 to 72 hours ahead: periods well within the in-sample. Then, the threshold value returning the smallest continuous ranked probability score (CRPS), defined in Section 3.4, is finally chosen at a particular lead time.

Figures 6 and 7 show how the forecast accuracy measured with CRPS varies depending on the model threshold value. As illustrated in Figure 6, for individual households, the model threshold that enhances forecast accuracy varies for lead time up to day ahead, which is of interest for our goal of improving market participants' strategies. For example, in the case of Korea, the optimal threshold was 0.20 for forecast lead times up to 9h-ahead. Beyond 9h-ahead, the optimal threshold started at 0.05 and reduced to 0.01 as the forecast lead time increased beyond 16h-ahead. For Ireland, the lowest CRPS is observed when the model threshold is 0.01 after 9h-ahead. However, as demonstrated in Figure 7, the most appropriate model threshold for each lead time is not substantially different for randomly aggregated demands. In the case of Korea, utilizing only the ARMA-GARCH model gives the greatest forecast accuracy at all lead times. For randomly aggregated demands of Ireland, using only the ARMA-GARCH model is the best choice until 15h-ahead, like in Korea. Beyond that, the threshold 0.01 continues to exhibit the lowest CRPS.

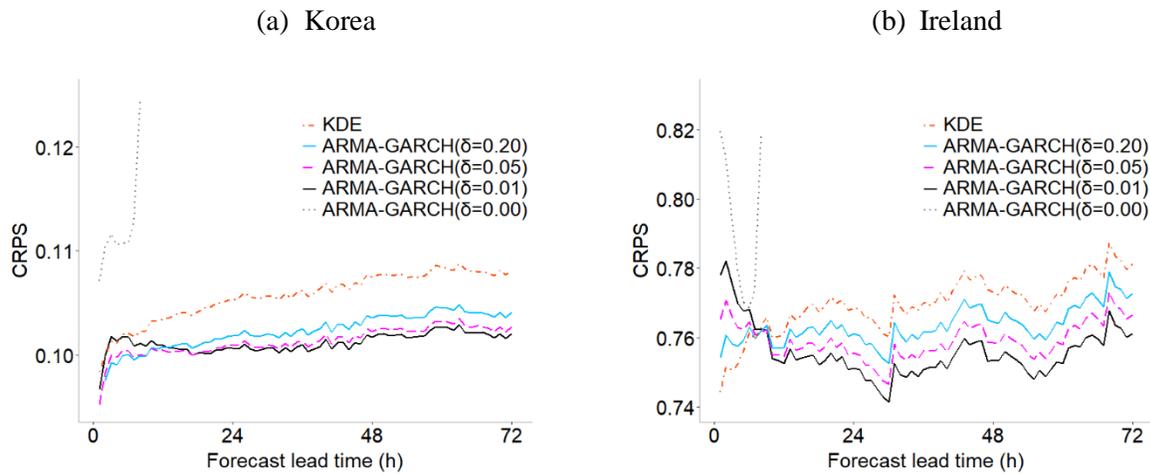

**Figure 6**. Probabilistic forecast accuracy of using various thresholds ($\delta$) for Goodness-of-fit of ARMA-GARCH for individual households.

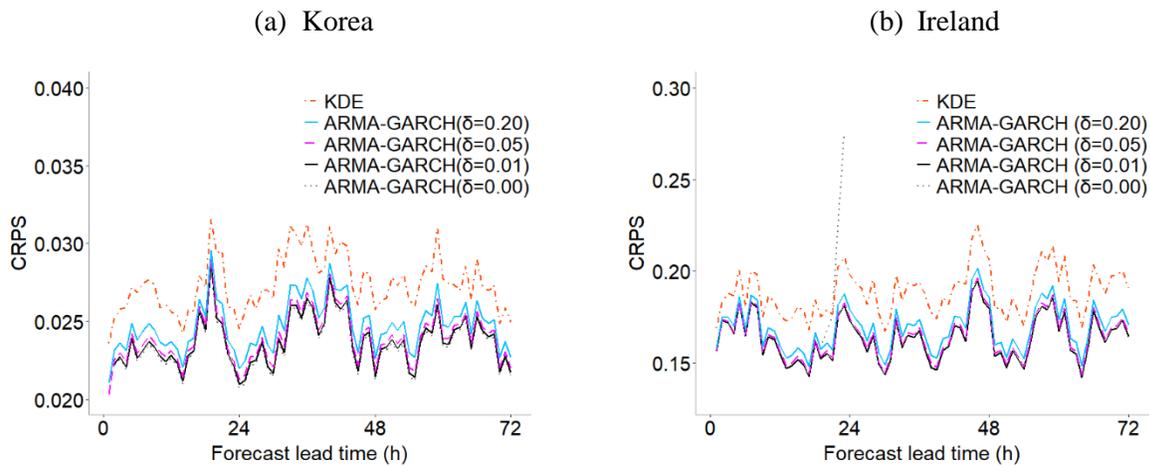

**Figure 7**. Probabilistic forecast accuracy of using various thresholds ($\delta$) for Goodness-of-fit of ARMA-GARCH for randomly aggregated demands.

*3.4. Evaluation criteria*

We use the continuous ranked probability score (CRPS), as described in Gneiting et al. [50], for the evaluation of the density forecast uncertainty. In probabilistic predictions of continuous variables, the score assesses the sharpness of the predictive distribution subject to calibration, as expressed in predictive densities or cumulative distribution functions (CDFs). Calibration assesses the congruence between the predictive distribution and the associated observations. Sharpness, a distinctive feature of the predictions, evaluates the concentrations of the predictive distributions [50]. The formal definition of this scoring method is described in Equation (4),

$$CRPS(F, y) = \int_{-\infty}^{\infty} \{F(x) - \mathbf{1}(x \geq y)\}^2 dy, \qquad (4)$$

where $F$ denotes the predictive CDFs, $y$ is the observation, and $\mathbf{1}$ is the Heaviside step function that returns 1 if the argument is positive or zero and 0 otherwise.

The CRPS can be expressed equivalently as in Equation (5), as demonstrated by Gneiting and Raftery [51],

$$CRPS(F, y) = E_F|X - y| - \frac{1}{2} E_F|X - X'|, \qquad (5)$$

where X and X′ are independent copies of a random variable with CDF $F$ and the finite first moment.

To evaluate the CRPS in this study, we apply Equation (5) and utilize predicted values and observations with trends and seasonal patterns eliminated, as outlined in Section 3.2. The trend and seasonal components are added to the predicted values to generate the final forecasts.

To evaluate the deterministic forecast accuracy, we also use the MAE, as defined by Equation (6), which compares the observations $y_t$ and deterministic forecasts $\hat{y}_t$ out of $n$ samples:

$$MAE = \frac{1}{n} \sum_{t=1}^{n} |\hat{y}_t - y_t| \qquad (6)$$

## 4. Demand aggregation

Load forecasting at lower levels of aggregation is becoming an increasingly critical issue as more DERs enter the electricity markets. Accurate load predictions may help with distribution network management in a variety of ways, including Peer-to-Peer (P2P), Demand Response (DR), Energy Management System (EMS), and Distributed Energy Resources (DER) integration [49]. The objective of our study is to propose a method for making optimum demand aggregation with minimum prediction errors for market players' strategic decisions. The size of an aggregation group could vary from one to hundreds.

To understand how the size of demand aggregation influences the variability of the aggregated time series, in Figure 8 we display the demand time series aggregated with four different numbers of randomly chosen households: (1) one, (2) ten, (3) one hundred and (4) two hundred. To ensure scale consistency, the aggregated demand value for each group is calculated as the average of its households' demand. The standard deviation values of the four time-series are (1) 0.101, (2) 0.066, (3) 0.050, and (4) 0.049 for the Korean data and (1) 0.692, (2) 0.586, (3) 0.526 and (4) 0.517 for the Irish data, indicating that the standard deviation tends to decrease greatly up to the aggregation of one hundred households; however, beyond this point, it declines slowly.

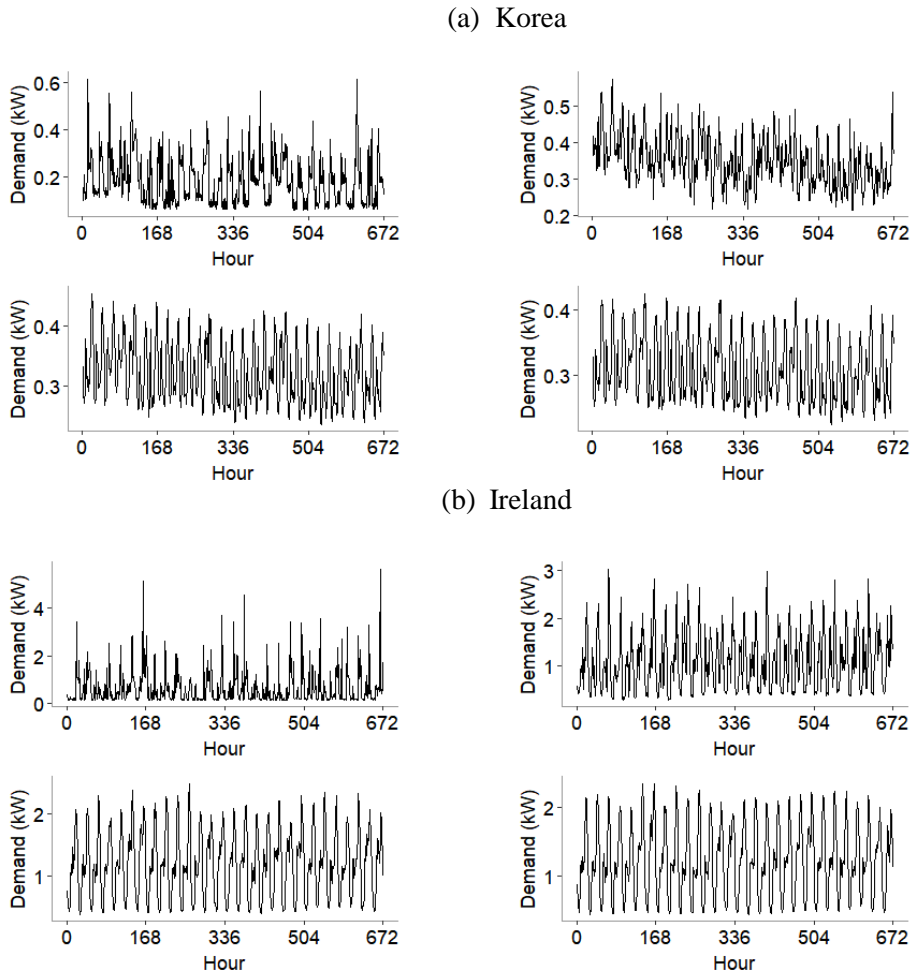

**Figure 8**. Example of demand time-series for one household (top-left), and the average of aggregated demand time-series for ten (top-right), one hundred (bottom-left), and two hundred (bottom-right) households chosen randomly during a period of four weeks.

For the four aggregation groups shown in Figure 8, we produced forecasts using the KDE model outlined in Section 3.2, and evaluated their forecast accuracy using the CRPS as shown in Figure 9. Similar to the patterns shown in Figure 8, in the time series averaged across households' load, a larger aggregate is anticipated to produce forecasts with lower forecast errors. Indeed, in Figure 9, there is a converging pattern that the increase in accuracy drops quickly beyond a random sample size of 100. This pattern is consistent at different forecast lead times, although it is sensible that the forecast accuracy decreases as the forecast lead time increases. This is supported by the power law relationship between aggregate size and forecast inaccuracy in previous studies [49,50].

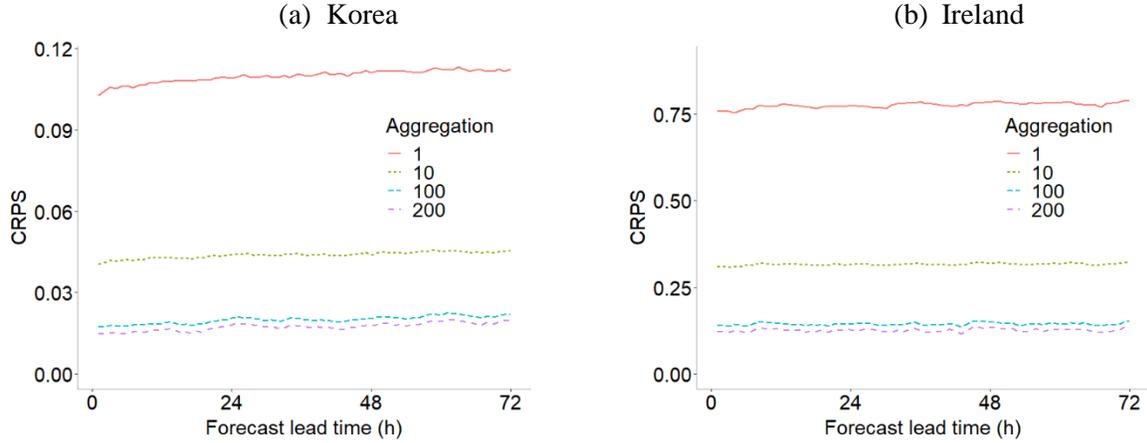

**Figure 9**. Forecast uncertainty for different sizes of random groups of households.

Figure 10 depicts the expected demand and CRPS of randomly selected groups at a certain lead time. Returning to Figure 1, the light grey dots represent numerous randomly formed groupings, whereas the dark grey points represent the best groupings in a range of randomly aggregated demand. In this study, we aim to present strategies for determining better aggregation for each demand range as outlined in Section 5. Ultimately, market participants in the electricity market can utilize the efficient frontier provided by the methods suggested in this paper and make better decisions for their strategies.

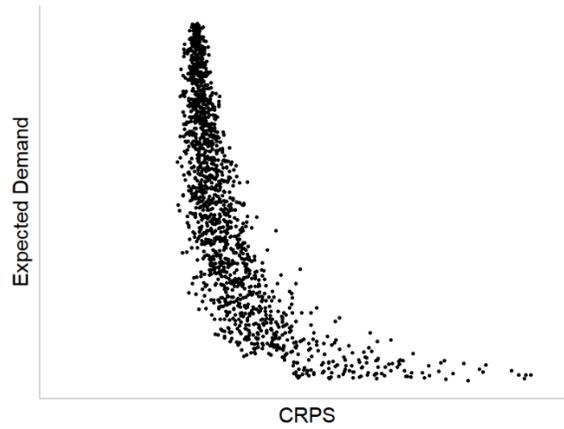

**Figure 10**. Expected demand and CRPS of random groups of households for the Korean data.

## 5. Forecast-based Portfolio Optimization

Portfolio optimization is a commonly used technique in finance and economics for creating a set of efficient portfolios with the highest possible return for any level of risk [22]. Outside of finance, the potential for portfolio optimization has been recognized in a variety of fields and is no different in the energy sector. Portfolio optimization may be applied in the electricity industry from both private (investors and managers) and public (regulators and planners) standpoints [25]. Faia et al. [51] present a new model for optimizing the portfolios of electricity market participation by maximizing their expected returns while minimizing risk (the volatility of electricity market price). Shahriari and Blumsack [23] build optimal portfolios of solar and wind in the Eastern United States to assess the capacity benefits of renewable portfolios in various areas of the electric grid.

In the last decade, much work has been done to extend classical portfolio optimization and make the modern portfolio theory more practical [52]. In the presence of constraints that restricts the amount of assets in a portfolio, as well as the proportions of the portfolio allocated to any given asset, Chang et al. [53] show that finding an efficient frontier becomes much more challenging, and offer three strategies based on genetic algorithms, tabu search and simulated annealing for discovering the constrained efficient frontier. Bonami and Lejeune [52] developed a method for finding exact solutions to portfolio optimization problems that accounts for both uncertainty in the estimate of the expected returns and real-life market limitations modeled with integer constraints. This work has two constraints. The first calls for an optimal demand portfolio that is comparable to a particular aggregated demand, while the second is a binary constraint that necessitates the selection of households. In this paper, we use a genetic algorithm to address these constraints, which has been shown to perform marginally better than tabu search and simulated annealing for the constrained portfolio optimization problems by Chang et al. [53]. The genetic algorithm developed by John Holland [54] uses generation-based optimization, modifying a population of trial solutions so that a new generation with better fitness is selected by applying crossover and mutation operations. With constrained portfolio optimization, we aim to develop an aggregated demand portfolio from various households that minimizes forecasting errors. We assume that market participants such as P2P operators, DR aggregators, and DSOs can select the best households to design the optimal portfolio for their strategies and that all the households are willing to participate.

Let $Y$ be a ($T \times N$) matrix containing $T$ demand observations and $N$ households. Then, at time $t$ we divide the range of the potential total demand of $N$ customers into $K$ equal-sized partitions to derive the optimal portfolio for each partition. $c_{(k,h)}$ and $c_{(k+1,h)}$ are the minimum and maximum values of the $k$-th partition at forecast lead time $h$. Then, the portfolio optimization in Equation (7) requires the definition of the objective function $g$, applied to $Y$:

$$min \ g\left(Y, v_t^{(k,h)}\right), \tag{7}$$

where $c_{(k,h)} < \widehat{Y}_{t+h} \cdot v_t^{(k,h)} < c_{(k+1,h)}$; $v$ is a ($N \times 1$) vector containing one or zero, which is the integrality constraint, indicating whether each of $N$ household will participate in the portfolio or not, respectively; and $\widehat{Y}_{t+h}$ is a vector of the energy demand forecasts of $N$ households at lead time $h$. The goal of the optimizer is the minimization of function $g$ by changing parameters such as the vector $v$. Hence, the vector $v$ is the final portfolio selection of households that minimizes the probabilistic forecast error at that specific forecast horizon $h$ and the demand at the $k$-th partition for the objective function $g$.

For the minimization, we used the genetic algorithm implemented by Mebane and Sekhon [55] in the R package, 'genoud'. In this work, the genetic algorithm is set to stop after 100 generations or after 10 consecutive generations without improvement in the objective function, whichever occurs first. For the objective function $g$, we propose three approaches: (1) forecast validated (FV), (2) seasonal residual (SR), and (3) seasonal similarity (SS), and describe them in detail in the following sub-sections:

(1) Forecast validated (FV): by minimizing forecast errors in the validation dataset.
(2) Seasonal residual (SR): by minimizing the standard deviation of its deseasonalized aggregated demand.
(3) Seasonal similarity (SS): by minimizing the deviation of the seasonal signals of customers in a portfolio.

*5.1. Forecast Validated*

The FV approach is the most straightforward of the three selection methods. The objective is the optimization of the group selection based on the forecast accuracy. Specifically, FV divides the in-sample data into a training matrix $Y_a$ and a test matrix $Y_b$. A density forecast $f$ is then produced using the training matrix, while the evaluation occurs against the test matrix as

$$g_{FV}\left(Y, v_t^{(k,h)}\right) = CRPS\left(f\left(Y, v_t^{(k,h)}\right)\right). \tag{8}$$

The objective function $g_{FV}$ returns the CRPS evaluation result of the density forecasts from the forecasting model $f$, defined in Section 3. We choose the optimal portfolio vector $v$ in the cross-validation period within in-sample for each forecast lead time $h$. For example, in the FV approach, a group created for 4h-ahead forecasts may not be optimal for 12h-ahead forecasts. Furthermore, one of the features of FV is that it allows households, even with highly volatile forecasts, to be included in the portfolio as long as the portfolio's forecast is accurate over the cross-validation period. For example, another household can exhibit an opposite but still highly volatile consumption behavior, and the two types of households can offset each other, producing less volatile consumption patterns.

*5.2. Seasonal Residual*

FV measures forecast uncertainty, but suffers from high computation time for probabilistic model estimation and density forecast evaluation in the cross-validation. Therefore, we introduce a relatively concise but effective approach with Seasonal Residual (SR). In the first step, the time-series of aggregated demand is created by multiplying $Y$ by the household selection vector $v_i$. SR then divides the aggregated demand into three components, (S)easonal, (T)rend and (R)emainder, a decomposition method detailed in Section 2.2 as formulated in the following:

$$STL(Y \cdot v_t^k) = S^{Y \cdot v_t^k} + T^{Y \cdot v_t^k} + R^{Y \cdot v_t^k}. \tag{9}$$

In Equation (10), the objective function $g_{SR}$ is defined as the relative standard deviation (RSD) of the remainder components from the aggregated demand:

$$g_{SR}(Y, v_t^k) = \widehat{RSD}\left(R^{Y \cdot v_t^k}\right), \tag{10}$$

where $\widehat{RSD}(\cdot)$ denotes the relative standard deviation, which is calculated as the sample standard deviation divided by its mean, and $R^{Y \cdot v_t^k}$ is a ($T \times 1$) vector containing the remainder components of the $k$-th aggregated demands. Minimizing $g_{SR}$ will choose the portfolio with the smallest standard deviation in the past. Another benefit of the model is that it can avoid expensive computation cost for each forecast lead time and validation required by FV.

The remainder (residual) component, which may represent a non-temperature-related element in electricity consumption, could help to increase overall forecasting accuracy [56]. For this reason, various approaches to investigating residual component modeling have been developed. Amara et al. [56] provide an adaptive non-parametric technique for extracting the behavior of the periodic part of the residual load.

*5.3. Seasonal Similarity*

The SS approach is a bi-objective optimization employing the objective ratio $r$. While FV and SR both utilize $v$ to multiply $Y$ and generate a single aggregated time series, SS uses $Y(v_t^k)$, which chooses a subset of the matrix $Y$, where the households are selected by $v$.

Similar to Equation (9), SS performs a seasonal decomposition but on each household level:

$$STL(Y(v_t^k)) = S(v_t^k) + T(v_t^k) + R(v_t^k). \qquad (11)$$

Then, Matrix $S(v_t^k)$, $T(v_t^k)$ and $R(v_t^k)$ consist of the seasonal, trend, and remainder time-series of each household selected with $v_t^k$, respectively. The SS approach minimizes the equation below by allowing the portfolio with minimal variations in both (1) the seasonal and trend and (2) the remainder components to be chosen, where their bi-objective weighting ratio is controlled by $r$:

$$g_{SS}(Y(v_t^k), r) = r^h \cdot mean\left(\widehat{RSD}(S(v_t^k) + T(v_t^k))\right) + (1 - r^h) \cdot mean\left(\widehat{RSD}(R(v_t^k))\right), \qquad (12)$$

where $mean(\widehat{RSD}(\cdot))$ indicates the average of the RSD values over different households. The goal of the optimizer of the objective function $g_{SS}$ is the minimization of Equation (12), by varying the vector $v$ and the weight $r$ within in-sample.

We allowed the weight $r$ varies with forecast lead time and summarized the CRPS evaluation results of the various weights assessed in Table 1. The weight $r$ that promotes predictability for the specified lead time shows very similar results for Korea and Ireland. In both countries, the density forecast accuracy of the SS approach was higher when the weight $r$ is 0.5 to 1.0, suggesting that combined households with more similar seasonal and trend components are preferable. Allowing different $r$ chosen for each lead time provided better accuracy than having the same $r$.

**Table 1**. SS optimization results for various configurations on bi-objective function weights in terms of CRPS (x100, kW).

| Country | SS weight $r$ | Average CRPS | | |
|---|---|---|---|---|
| | | **4h-ahead** | **12h-ahead** | **24h-ahead** |
| **Korea** | 1.0 | 1.01 | 0.99 | 0.93 |
| | 0.8 | 0.98 | 1.04 | 0.90 |
| | 0.5 | 1.00 | 1.02 | 0.91 |
| | 0.2 | 1.05 | 0.99 | 0.92 |
| | 0.0 | 1.06 | 1.07 | 0.97 |
| **Ireland** | 1.0 | 6.98 | 7.31 | 8.18 |
| | 0.8 | 7.01 | 6.90 | 8.48 |
| | 0.5 | 7.39 | 6.74 | 8.32 |
| | 0.2 | 7.31 | 6.85 | 8.79 |
| | 0.0 | 7.57 | 7.09 | 8.26 |

In brief, the SS approach determines a combination that minimizes the variance of an individual household's seasonal, trend, and remainder components at a given time, whereas the SR approach finds a combination that minimizes the variation of the residual part of aggregated households' total power consumption over time. Furthermore, as compared to SR, SS necessitates an additional step to identify the optimal weight $r$ in advance.

## 6. Results and Discussion

### 6.1. Empirical Results

This section compares the forecasting performance of the three proposed approaches that identify demand groups that would lower the uncertainty, as described in Section 5: FV for minimizing the forecast error, SR for minimizing the standard deviation of the residuals of deseasonalized aggregated demand, and SS for minimizing the household-level deviation of the seasonal signals and the residuals. As shown in Section 4, there is no substantial improvement in prediction accuracy for the aggregation of more than 100 households; thus, we designed portfolio optimization for a capacity of up to 100 households. We focus on day-ahead and intraday forecasts, which are crucial to the decisions in the electricity market. For example, accurate short-term load forecasting is certainly helpful for P2P participants' trading strategies, DSOs' grid flexibility utilization, and DR aggregators' flexibility services for day-ahead and real-time markets.

Figures 11 to 16 present the portfolios optimized for 4-hour, 12-hour, and day-ahead forecasts for Korean and Irish households using CRPS and MAE as evaluation criteria. As described in Section 2.1, the average of 48 rolling evaluations with a fixed prime number interval to avoid intraday seasonal bias was used in the figures. The model threshold $\delta$ for the choice between ARMA-GARCH and KDE and the ratio weight $r$ of the SS approach was set differently with forecast lead times, as shown in Sections 3.3 and 5.3.

Figures 11-16 demonstrate that the proposed methods - FV, SR and SS – consistently improved the forecast accuracy in terms of CRPS and MAE as expected demand decreases, while the random approach combination approach yields the opposite trend. Notably, the three proposed methods outperformed the random combination in terms of CRPS and MAE when the expected demand is low. This pattern is consistent in Table 2 where the forecast accuracy was averaged over different demand sizes and forecast lead times: 4h, 12h, and 24h-ahead. When comparing the three proposed methods, FV performed the best in both Korea and Ireland, followed by SR and then SS.

**Table 2**. Average CRPS (x100, kW) and MAE (x100, kW) of the proposed approaches for 4h, 12h, and 24h-ahead forecasts.

|  | Korea | | Ireland | |
| --- | --- | --- | --- | --- |
| **Approach** | **CRPS** | **MAE** | **CRPS** | **MAE** |
| **Random** | 2.2 | 2.9 | 18.0 | 23.7 |
| **FV** | 0.9 | 1.1 | 7.3 | 9.6 |
| **SR** | 1.0 | 1.3 | 8.4 | 10.9 |
| **SS** | 1.0 | 1.3 | 8.5 | 11.0 |

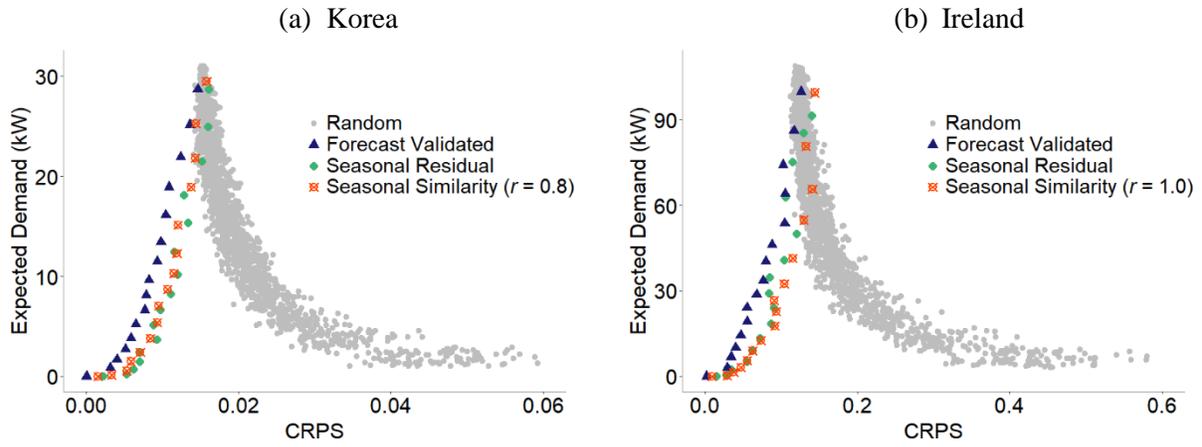

**Figure 11**. Portfolio optimization of 4h-ahead forecasts using CRPS.

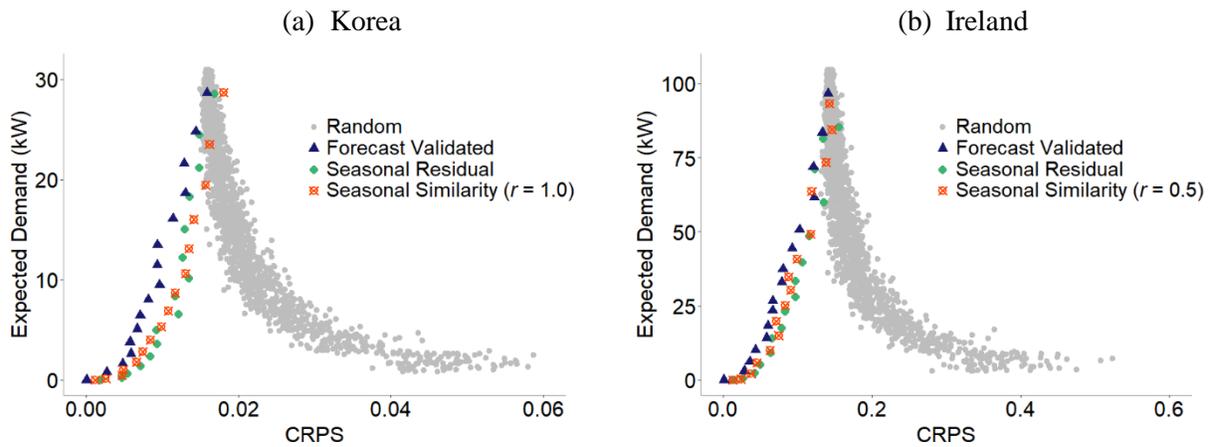

**Figure 12**. Portfolio optimization of 12h-ahead forecasts using CRPS.

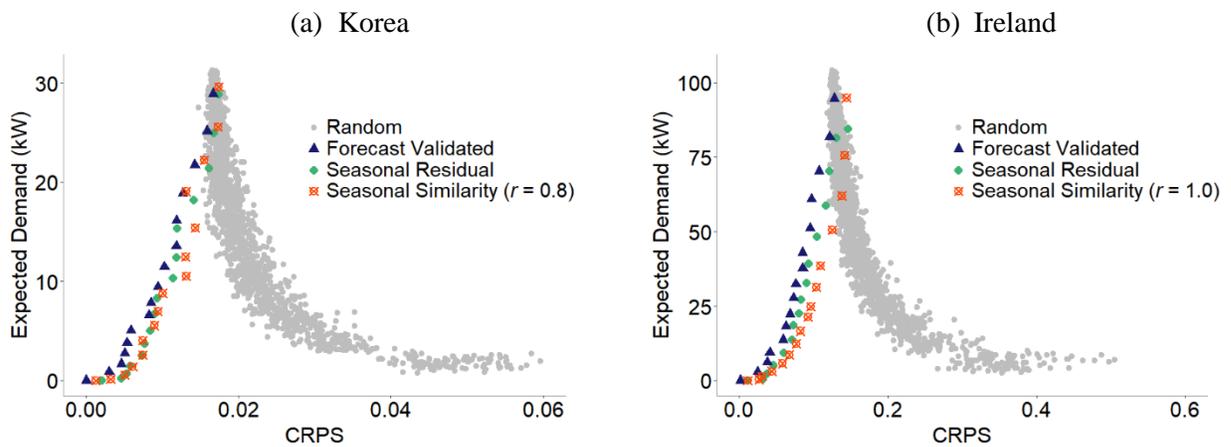

**Figure 13**. Portfolio optimization of 24h-ahead forecasts using CRPS.

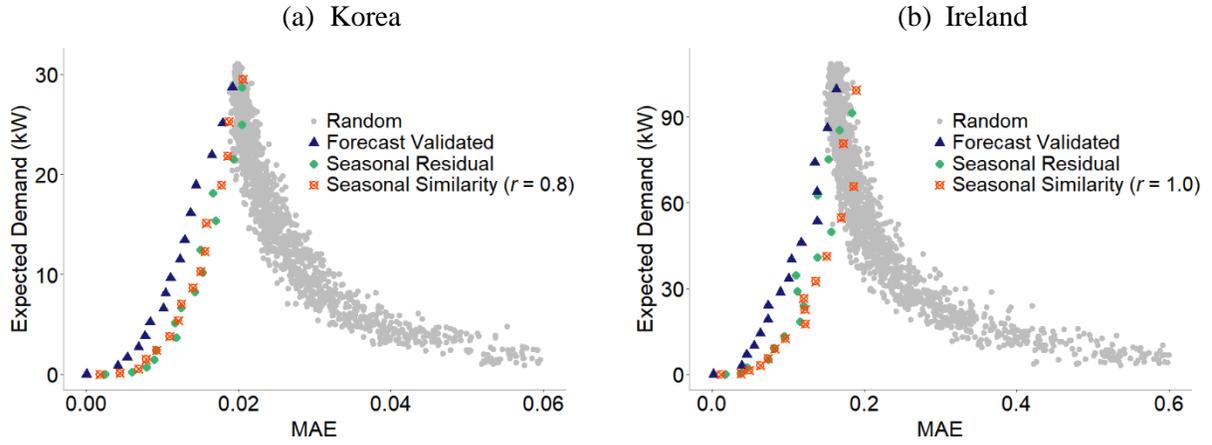

**Figure 14**. Portfolio optimization of 4h-ahead forecasts using MAE.

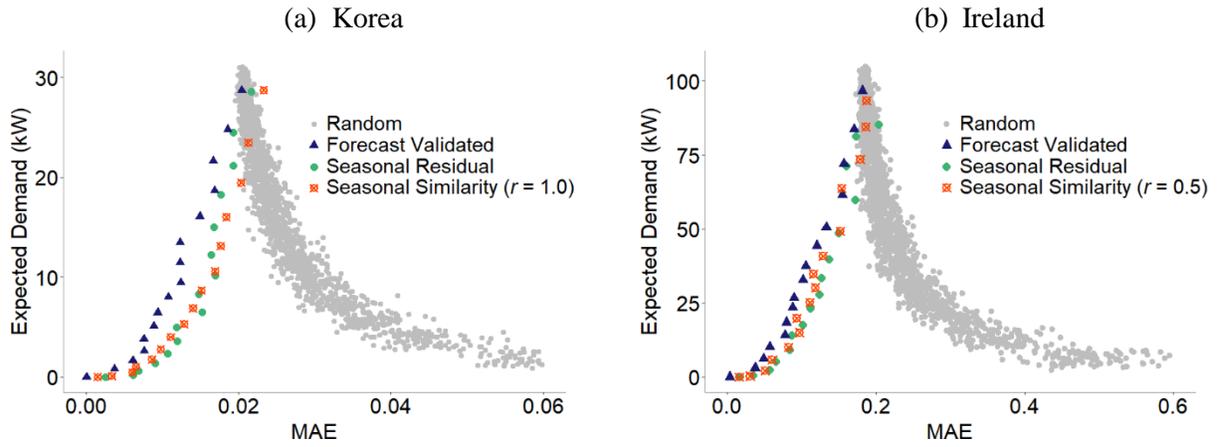

**Figure 15**. Portfolio optimization of 12h-ahead forecasts using MAE.

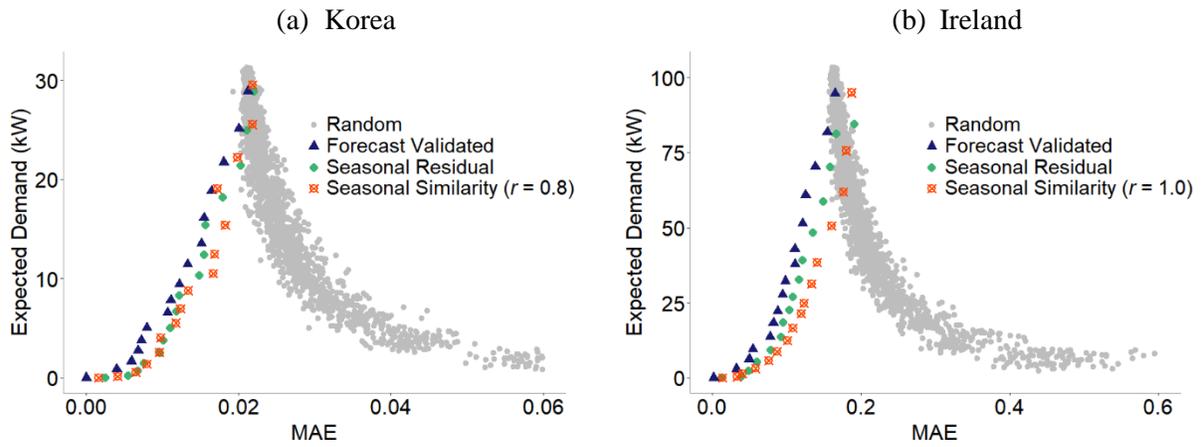

**Figure 16**. Portfolio optimization of 24h-ahead forecasts using MAE.

Figure 17 further illustrates the improvements in CRPS and MAE of the proposed methods over the random approach by lead time in Korea and Ireland. All of the proposed methods reduced the CRPS and MAE by an average of 51-62% in both cases when compared to the random approach. The FV approach showed the most significant performance increase in all three lead times in Korea and Ireland. The SR method marginally outperformed the SS method in the 24h-ahead forecast in Korea and the 4h-ahead and 24h-ahead forecasts in Ireland, whereas the SS method slightly outperformed SR in the 4h-ahead and 12h-ahead forecasts in Korea as well as the 12h-ahead forecast in Ireland.

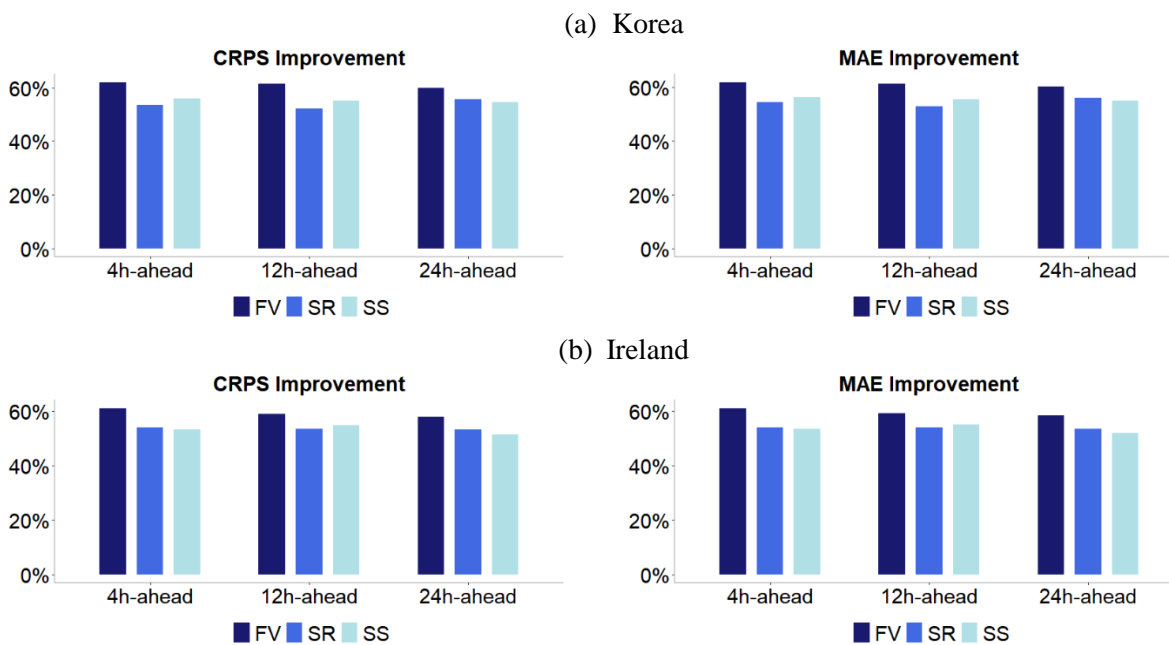

**Figure 17**. Performance enhancement of the proposed methods relative to the random by lead time

In summary, FV outperformed all the other methods in both Korea and Ireland across all aggregated demand sizes and lead times, but its slight improvement over SR and SS came at the cost of high computational expense and the need for parameter optimization for different forecast lead times. In contrast, the variability estimation approaches, such as SR and SS, require substantially less computation cost as they do not require choosing a different group for different forecast lead times, but instead compose the group only with a high likelihood of minimal standard deviation from a deterministic seasonal decomposition. Indeed, in our experiments using an 8-core i7-9700F processor to estimate the model parameters, the FV optimization method took four hours, while the SR and SS approaches took less than five minutes.

Based on Table 2 and Figure 17, SR and SS methods performed similarly, but SR was more effective as it required less computation cost while maintaining a similar forecasting accuracy. SS takes one additional step in advance to set the weight $r$.

Finally, we explored whether a relaxation of the binary constraint would provide different results by allowing the SR function for 4h-ahead forecasts to have portions of each household in the portfolio selection. This relaxation offered a slight benefit in terms of CRPS as shown in Figures 18 and 19. The CRPS (x100) of the relaxed version of SR was 1.02 and 8.17 for Korea and Ireland, respectively, which was slightly lower than the CRPS (x100) value of 1.03 and 8.22 for the unrelaxed version.

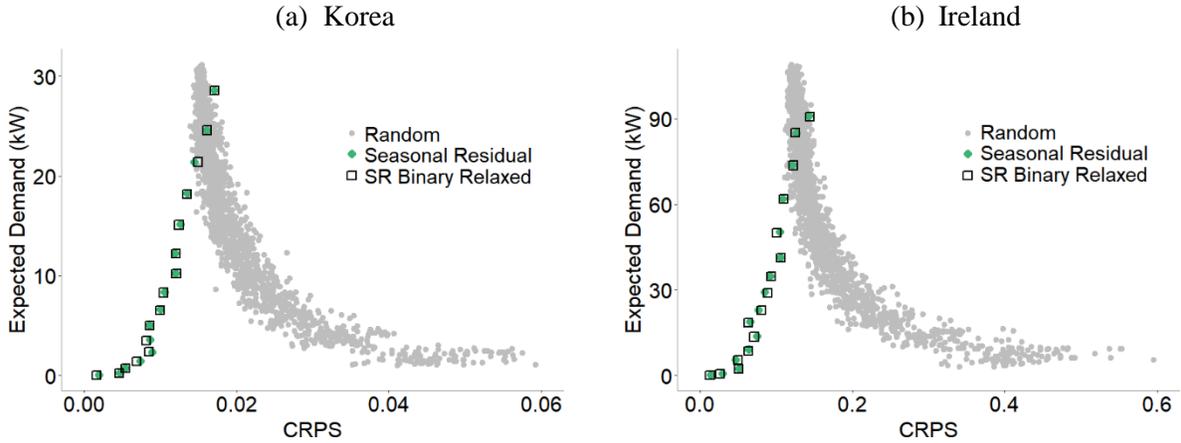

**Figure 18**. Portfolio optimization relaxing the binary constraint on the 4h-ahead forecast using CRPS.

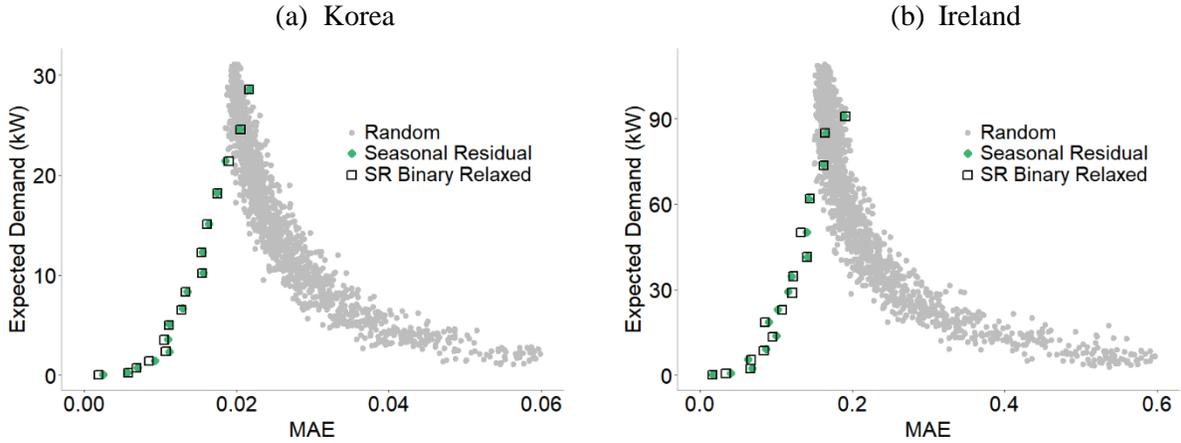

**Figure 19**. Portfolio optimization relaxing the binary constraint on the 4h-ahead forecast using MAE.

*6.2. Practical Implications*

The proposed methods in this study aim to increase forecasting accuracy by optimally combining the electricity consumption of residential households with high uncertainty, and can have many practical implications for the electricity market. With the spread of DERs, P2P energy sharing, distribution grid balance, and flexibility service of DERs including DR are realized in many countries. The SR approach particularly can provide both prediction accuracy and computational efficiency in the load forecasting at lower aggregation levels, and help make strategic decisions in the decentralized market. For example, accurate day-ahead and intra-day forecasts are essential for the trading strategies of P2P participants, the grid flexibility utilization of DSOs, and the flexibility services of DR aggregators.

First, consumers and prosumers in P2P energy sharing can make day-ahead or intraday scheduling of DERs more accurately using the proposed methods, which would help the participants to choose the most efficient strategy for their capacities of PV, EV, and ESS, and to invest in, manage and participate in P2P sharing. Ultimately, this will encourage P2P participants to reduce their energy expenditures, generate income, and facilitate P2P energy sharing markets.

Second, our suggested method can benefit DSOs in managing local power distribution sources efficiently for P2P energy sharing and DERs, including DR as a flexibility resource to manage the distribution grid, by providing accurate demand forecasts at the low aggregation level. Additionally, DSOs can participate in the balancing market with DERs to help balance the main grid, which requires cooperation with transmission system operators (TSOs), and this can also be facilitated by our proposed method.

Third, the reduced uncertainty in the forecast based on the proposed methods can also benefit DR aggregators or utilities to design effective DR programs. One of the primary challenges for DR deployment is establishing a DR baseline, which is an estimate of how much electricity a customer would use if the DR program were not in place. Accurate prediction of the baseline load of optimally selected customers using the suggested approaches will help to estimate the available aggregated DR capacity in the day-ahead and intraday markets and engage consumers more in DR programs by offering fair remuneration to DR participants.

In brief, market participants may enhance load forecast accuracy and make more effective decisions by using the proposed methods, depending on the lead time and demand group size required. The diffusion of DERs through this approach can contribute to grid balancing as a flexibility resource and, ultimately, help with carbon neutrality.

## 7. Conclusion

As the electricity market becomes increasingly decentralized for lower carbon use, the demand for short-term probabilistic load forecasting at low aggregation levels has increased for various market participants' decision making. In this study, we present a novel approach for constructing an aggregated demand portfolio with the lowest probabilistic prediction errors for a given energy demand, which can be used as a foundation for the strategic decision making of market participants on topics such as effective P2P operation, sustainable distribution systems and DR facilitation. Although multiple methods, such as forecast model development, have recently been investigated to enhance forecast accuracy at lower aggregation levels, research integrating portfolio optimization theory with consumers' demand portfolio optimization to minimize prediction error is still lacking.

In this paper, we propose three different objective functions to create a portfolio of residential households' demand: Forecast Validated, Seasonal Residual, and Seasonal Similarity. We evaluate the density forecasts of group demand built by these three approaches to determine the best method. The ARMA-GARCH model supported by an unconditional KDE model was employed for probabilistic load forecasting. The CRPS and MAE are used to evaluate density and deterministic forecasts.

The results show that all three functions improve forecast accuracy compared to random portfolios. In terms of forecast performance and computational cost, the SR based household portfolios for Korea and Ireland are the best overall. Market participants can make fast and accurate decisions using the proposed methods by enhancing the load prediction accuracy at low levels of aggregation, which are crucial for peer-to-peer sharing, distribution grid balancing, and demand response.

Additionally, it is practical to develop optimal aggregated demands to increase forecast accuracy while accounting for locational constraints. Therefore, in future work, we may evaluate the suggested technique using data with locational information.

**Acknowledgements**

The research was supported by the EPSRC grant (EP/N03466X/1), a National Research Foundation of Korea grant and the KISTI Super Computing Center. We are grateful for the insight comments of participants at the International Symposium on Forecasting, 2018.

**Author Contributions**

**Conceptualization**: Jooyoung Jeon, Ran Li, Fotios Petropoulos, Jungyeon Park, Estêvão Alvarenga.
**Data curation**: Jungyeon Park, Jooyoung Jeon, Estêvão Alvarenga.
**Formal analysis**: Jungyeon Park.
**Funding acquisition**: Jooyoung Jeon.
**Investigation**: Jooyoung Jeon.
**Methodology**: Jooyoung Jeon, Ran Li, Fotios Petropoulos.
**Project administration**: Jooyoung Jeon.
**Software**: Jungyeon Park, Estêvão Alvarenga, Hokyun Kim.
**Validation**: Jungyeon Park, Hokyun Kim, Kwangwon Ahn.
**Visualization**: Jungyeon Park.
**Writing – original draft**: Jungyeon Park, Estêvão Alvarenga.
**Writing – review & editing**: Jooyoung Jeon, Ran Li, Fotios Petropoulos, Kwangwon Ahn